\documentclass[aip,pop,
reprint%
]{revtex4-1}
\usepackage{graphicx}
\usepackage[utf8]{inputenc}
\usepackage[T1]{fontenc}
\usepackage{mathtools}
\usepackage{bm}
\usepackage{amsmath}
\usepackage{amssymb}
\usepackage{mathptmx}
\usepackage{xcolor}
\usepackage{comment}
\usepackage{units}
\usepackage[bookmarks=true,bookmarksnumbered=false,bookmarksopen=false,
 breaklinks=false,pdfborder={0 0 0},pdfborderstyle={},backref=false,colorlinks=true]
 {hyperref}
\hypersetup{pdftitle={Single mode wave decay},
 linkcolor=blue, citecolor=blue}
 
\usepackage{etoolbox}

\makeatletter
\def\@email#1#2{%
 \endgroup
 \patchcmd{\titleblock@produce}
  {\frontmatter@RRAPformat}
  {\frontmatter@RRAPformat{\produce@RRAP{*#1\href{mailto:#2}{#2}}}\frontmatter@RRAPformat}
  {}{}
}%
\makeatother

\begin{document}


\title{Single Mode Wave Decay} 



\author{Guilherme T. Irum\'e}%
 \affiliation{Instituto de F\'{\i}sica, Universidade Federal do Rio Grande do Sul, CP 15051, 91501-970, Porto Alegre, RS, Brazil}%
 \email{guilherme.irume@ufrgs.br}

\author{Joel Pavan}
\affiliation{Instituto de F\'{\i}sica e Matem\'atica, Universidade Federal de Pelotas, CP 354, 96010-900, Pelotas, RS, Brazil}%

\author{Rudi Gaelzer}
\affiliation{Instituto de F\'{\i}sica, Universidade Federal do Rio Grande do Sul, CP 15051, 91501-970, Porto Alegre, RS, Brazil}%



\date{\today}

\begin{abstract}
The usual approach on electrostatic wave decay process for a weak beam-plasma system considers two different wave modes interplaying, the Langmuir and ion-sound mode.
In the present paper, a single mode approach is shown to be feasible for conditions where the respective dispersion relations undergo topological changes. Numerical solutions for the dispersion relation of a beam-plasma system are presented, supporting the modeling of an analytic dispersion relation of a single wave mode. This wave mode is accounted for
in the kinetic equations for particles and waves, which rule the evolution of the system. The results are compared against the two-wave mode approach using Langmuir and ion-sound waves, within the context of weak turbulence theory. It is found that the single mode approach can account for the basic features of particles and waves, since the single mode exhibits both a region of low and high frequency which ultimately play the roles of ion-sound and Langmuir modes, respectively.
\end{abstract}


\maketitle 

\section{Introduction}

Explosive events that energize and accelerate particles are ubiquitous
in space and astrophysical environments. Of particular importance
are phenomena that generate beams of energetic electrons that will
propagate through a, usually denser, background plasma composed also
of electrons and ions. This kind of setup is usually called a beam-plasma
system and it has been observed in solar flares and coronal mass ejections,\citep{Hudson11/01,White+11/07,Chen+13/01,ReidRatcliffe14/07,ReidKontar18/06,Badman+22/10,LorfingReid23/04,Lorfing+23/12,Khoo+24/03}
the foreshock region of Earth's magnetosphere\citep{Soucek+19/04}
and in the auroral region,\citep{Akbari+21/03} among other space
environments.

The excess of free energy contained in the electronic beam frequently
converts to electrostatic and/or electromagnetic waves through a series
of physical processes. In fact, in many regions of the heliosphere
and planetary magnetospheres the remote observation of free-propagating
electromagnetic waves was the first and only possible evidence of
the energetic beams. Such was the case of the solar corona. For decades,
before \emph{in situ} observations were possible thanks to the recently-launched
Parker Solar Probe and Solar Orbiter, the properties of particle beams
originating from the solar chromosphere and expanding into the solar
corona were inferred by wave phenomena such as solar Type III and
X-ray emissions\@.\citep{LorfingReid23/04,Badman+22/10,ReidKontar17/02,ReidKontar18/06,ReidRatcliffe14/07,Sauer+19/01,White+11/07,Hudson11/01,Lorfing+23/12}
The same happens with radio emissions from the auroral regions of
other planets of the solar system, such as Jupiter and Saturn, which
are belived to be caused by energetic electron fluxes accelerated
along the magnetic field lines, sometimes in combination with other
free-energy sources such as loss-cone or ring distributions.\citep{Menietti10/04,Mutel+10/10,Vorgul+11/05,Menietti+11/12,Melrose17/12,Lysak+23/01}

Regarding a pure beam-plasma system, one of the most accepted physical
mechanisms whereby free energy from the electronic beam is ultimately
converted to electromagnetic waves is the \emph{plasma emission} process,
which occurs in a series of steps, beginning by the direct, linear
conversion of particle kinetic energy into longitudinal Langmuir and
ion-sound waves, followed by nonlinear mechanisms that couple transverse
waves with the longitudinal waves, allowing the former to grow from
the enhanced level of the latter. A solid evidence for this mechanism
is the frequent \emph{in situ} observation of enhanced Langmuir waves
associated with particle beams.\citep{Soucek+05//08,Pisa+16/08,ReidKontar17/02,Soucek+19/04,Gomez-Herrero+21/12,LorfingReid23/04,Lorfing+23/12}

The subsequent nonlinear mechanisms that convert Langmuir waves into
other normal modes are usually assumed to occur in a time scale larger
than the quasilinear diffusion time. One recently proposed mechanism
starts with the parametric decay of Langmuir waves into other longitudinal
modes such as the electron- and ion-acoustic waves, followed by conversion
to electromagnetic waves.\citep{Soucek+05//08,Sauer+17/07,Sauer+19/01}

Another frequently proposed mechanism is based in the weak turbulence
theory derived from a perturbative approach to the Klimontovich-Dupree
formalism.\citep{Sitenko82,Yoon19} According to the theory, the particle
velocity distribution functions and the spectral intensities of the
existing normal modes of oscillation evolve in time according to kinetic
equations that include the processes occurring during the quasilinear
relaxation phase, followed by the nonlinear three-wave decay and nonlinear
wave-particle scattering that occur in a longer time scale. The weak
turbulence theory in particular, and the plasma emission hypothesis
in general, have been developed and compared with numerical simulations
by several contributions along the years, of which Refs. \onlinecite{Yoon00/12,ZiebellGaelzerYoon01/09,YoonGaelzer02/10,YoonGaelzer02/11,Yoon+03/02,Gaelzer+03/02,Umeda+03/02,Yoon05/04,YoonRheeRyu05/11,ZiebellGaelzerYoon08/03,Gaelzer+08/04,Yoon10/11,Pavan+11/04,Ziebell+15/06,Ziebell+16/02,KrafftVolokitin16/04,Lee+19/01,Henri+19/03,Chen+22/01}
is but a sample of recent contributions.

In the context of beam-plasma systems and weak turbulence theory, 
the self-consistent evolution of longitudinal modes in a field-free plasma
is usually described in terms of two interplaying modes, namely, Langmuir and ion-sound modes\@.\cite{Cairns89/01,Gaelzer+03/02,YoonGaelzer02/11,YoonRheeRyu05/06} Specifically, the wave-wave process accounted is the 3-wave decay involving both modes. Despite that, 3-wave decay involving ion-sound mode only and 4-wave decay involving Langmuir mode only, for instance, are feasible. \cite{Davidson68/12,Davidson69/01,Lvov+97/07} %

Wave decay processes play an important role in regulating wave spectra and particle velocity distributions in plasma systems. In particular, beam-plasma systems are shown to be significantly influenced by wave decay processes. The higher the energy carried by the beam, the more relevant the decay processes are shown to be, since such processes usually take place as higher order nonlinear contributions.\cite{Gurnett+93/04,Yoon00/12,m86,BohmGross49/06,b64} Henceforth, it is emphasized the meaning of wave decay as a wave-wave process, that is to say, a process involving waves only, not particles. This distinction is called for since waves can decay intermediated by particles, as in the so-called nonlinear wave scattering process.

The present work addresses a beam-plasma system, aiming to describe the system in terms of a single wave mode, taking into account the quasilinear interaction with particles and the nonlinear interaction among waves, through the three-wave process, and compare against the usual approach that accounts for Langmuir and ion-sound modes.\cite{Ziebell+15/06,Ziebell+16/02,Lee+19/01,Pisa+16/08,Akbari+21/03,Zhou+20/03,Sun+22/09a,Gaelzer+08/04,Yoon19}



Although not usual in the study of the beam-plasma process (see, \emph{e.g.}, Ref. \onlinecite{Yoon19} for a list of references), a single-mode approach is by no means uncommon.  For instance, the nonlinear decay of Alfvén waves into a pair of other Alfvén or related waves in conditions pertaining to the solar wind plasma has been studied within the frameworks of MHD\cite{DelZanna+01/02,DelZanna01/07} and hybrid\cite{Matteini+10/10} simulations.
Three-wave decay processes among kinetic Alfvén waves were also studied using either kinetic theory\cite{Voitenko98/10} or a two-fluid formalism.\cite{VoitenkoGoossens05/07}
However, the description of decay processes in a beam-plasma system using the usual electrostatic normal modes needs the interaction of different waves in order to match the three-wave decay condition.  On the other hand, the present single-mode description employs a modified-beam mode that has a frequency range large enough to carry out the decay process.

The plan of the paper is as follows. In section \ref{sec:II}, numerical solutions of the electrostatic dispersion relation of a beam-plasma system are analyzed, providing support for the single mode modeling. In section \ref{sec:III}, the equations which establish the temporal evolution of particles and waves are presented, and the simulation results obtained both from the usual two-mode approach and single mode approach are compared.
Finally, in section \ref{sec:IV} we present our final remarks.
\section{Dispersion relation}
\label{sec:II}

For a system described by non-equilibrium (shifted Maxwellians) distribution functions, as occurs in the beam-plasma system, the dispersion relation for Langmuir waves (Bohm-Gross) $\omega_L^2=\omega_{pe}^2 +3v_{te}^2k^2$ and the dispersion relation for ion-sound waves $\omega_S^2=(c_S^2k^2)/(1+\lambda_{De}^2k^2)$ are often assumed to remain valid.
Here, $\omega_{pe} =\sqrt{4\pi n_0e^2/m_e}$ is the electron plasma frequency, 
$v_{te} = \sqrt{T_e/m_e}$ the thermal velocity of electrons, and
$c_S = \sqrt{T_e/m_i}$ the ion-sound speed. 
In the above, $n_0$ is the total plasma density, $m_e$ and $e$ the mass and charge of the electron, respectively, $T_e$ is the electron temperature in energy units,
$m_i$ is the ion mass and $\lambda_{De} = \sqrt{T_e/4\pi n_0 e^2}$ the Debye length. In this case, it is hypothesized that deviations from the distribution in relation to the (non shifted) Maxwellian will not significantly alter the dispersion of normal plasma oscillation modes, and their contributions are considered only in the absorption/emission coefficient.

However, more careful studies on the effect of the electron beam, not only on the absorption/emission coefficient but also on the dispersion relation of the modes,  have shown that in the case of a beam-plasma system the Langmuir and ion-sound dispersion relations are only valid under very restrictive conditions. \cite{Cairns89/01, ONeilMalmberg68/08, ThurgoodTsiklauri15/12,Soucek+19/04}

O'Neil and Malmberg \cite{ONeilMalmberg68/08} deduced that the Bohm-Gross dispersion relation is only valid when the %
scaled thermal spread 
%
\begin{equation*}
    s = \frac{v_{tb}}{v_{b}} \left(\frac{2n_0}{n_b}\right)^{1/3}
\end{equation*}
satisfies, when both the core and the beam distributions are Maxwellian, $s \gtrsim 1.47$. %
In the above expression $v_{tb} = \sqrt{T_b/m_e}$ is the thermal dispersion of beam particle velocities, $v_b$ the average beam velocity, and $n_b$ the beam density.

A more careful analysis developed by Cairns and Fung\cite{Cairns89/01,CairnsFung88/07}, resulted in even more restrictive conditions for the validity of this approach. 
The results obtained by the Authors
indicate that the usual weak turbulence theory must be modified to describe the behavior of a beam-plasma system when the beam is of higher intensity, meaning higher density or velocity. The necessary modifications involve the consideration of the oscillation modes relevant to the case of intense beams, as well as the use of the respective dispersion relations and coupling terms of waves in order to circumvent the inconsistency arising when considering beams of higher intensity.\cite{Kainer+72/12,Sauer+19/01}

Nevertheless, we hasten to emphasize that the developments presented in this paper aim solely to indicate a transition of regimes found when the beam energy is increased. This is noteworthy since the parameters involved in such a transition are on the edge of validity of the weak turbulence theory, used as the theoretical framework.

The dispersion equation employed in this work is obtained from the Vlasov-Poisson system of equations and determines the dispersion of longitudinal oscillations in a plasma without external fields applied. This equation is written as \cite{Cairns89/01}
\begin{equation}
D(k,\omega) = 1 + \sum_{\alpha = e, i, b} \frac{\omega^2_{p\alpha}}{k^2} \int d\mathbf{v} \frac{\mathbf{k} \cdot \partial {f}_{\alpha}({\mathbf{v}})/\partial \mathbf{v}}{\omega - \mathbf{k} \cdot \mathbf{v}} = 0 \,,
\end{equation}
where $f_{\alpha}(\mathbf{v})$ is the velocity distribution function and $\omega_{p\alpha} = \sqrt{4\pi n_\alpha q_\alpha^2/m_\alpha}$ the plasma frequency, being the usual quantities identified for each species $\alpha$ as \textit{i}, \textit{e}, \textit{b}, standing for thermal ions and electrons, and electron beam, respectively. For a plasma system composed by three maxwellian populations of ions, thermal electrons and an electron beam, in one-dimensional space for velocity and wave number, the initial distribution function for species $\alpha$ is 
\begin{eqnarray*}
&& f_{\alpha}(v) = \frac{1}{\sqrt{\pi}v_{t\alpha}} \exp{\frac{{-(v-v_{\alpha}})^2}{v_{t\alpha}^2}}\,,
\end{eqnarray*}
where $v_{t\alpha} = \sqrt{T_\alpha/m_\alpha}$ and $v_{\alpha}$ are the thermal speed and drift velocity of species $\alpha$, respectively. 

With these definitions, the dispersion equation for electrostatic waves is well-known and is given by \cite{Cairns89/01}
\begin{multline}
1- \frac{\omega^2_{pi}}{k^2 v^2_{ti}} Z' \left(\frac{\omega}{k v_{ti}}\right) - \frac{\omega^2_{pe}}{k^2 v^2_{te}} Z' \left(\frac{\omega +  k v_e}{k v_{te}}\right)
\\ 
- \frac{\omega^2_{pb}}{k^2 v^2_{tb}} Z' \left( \frac{\omega - k v_b}{k v_{tb}} \right) = 0 \,,
\label{eq:3}
\end{multline}
where $n_i = n_e + n_b = n_0$ and $v_e = {n_bv_b}/{n_e}$,
satisfying the conditions of charge neutrality and zero current, respectively. The relation between drift velocities implies that the analysis is performed in a reference frame at rest with the ions. Finally, $Z'{(\zeta)}$ is the derivative of the  Fried and Conte function.\cite{Robinson89/11}

When it is assumed that the beam is very tenuous, which is tantamount to a low energy beam, and consequently the system is weakly unstable, an analytic solution is possible for the dispersion equation, and results in the well-known Bohm-Gross dispersion relation for the high frequency range (the Langmuir mode), and the ion-sound mode for the low frequency range. However, for the case of high energy beams a numerical solution is called for.

It is possible to analyze and classify the numerical solutions based on the criteria presented by Cairns\@ \cite{Cairns89/01}%
    \begin{equation*}\label{P-definition}
        P=2^{1/3}/s,
    \end{equation*}
in which for $P<1$ the instability is kinetic and for $P>1$ the instability is reactive. Additionally, for $0.2 \lesssim s \lesssim 1.4$ $(0.9 \lesssim P \lesssim 6.3)$,
the instability is resonant but occurs along a propagation mode called beam-modified mode, which is characterized by a linear relation of type $\omega \approx kv_b$  for small $k$\@. For $s \gtrsim 1.5$ $(P \lesssim 0.84)$
the instability is strongly resonant and occurs along an asymptotically approaching mode of the Langmuir mode. %
Thus, it is concluded that systems with more intense beam will have the value of $P$ increased and the value of $s$ decreased.

\begin{figure*}
    \includegraphics[width=\textwidth]{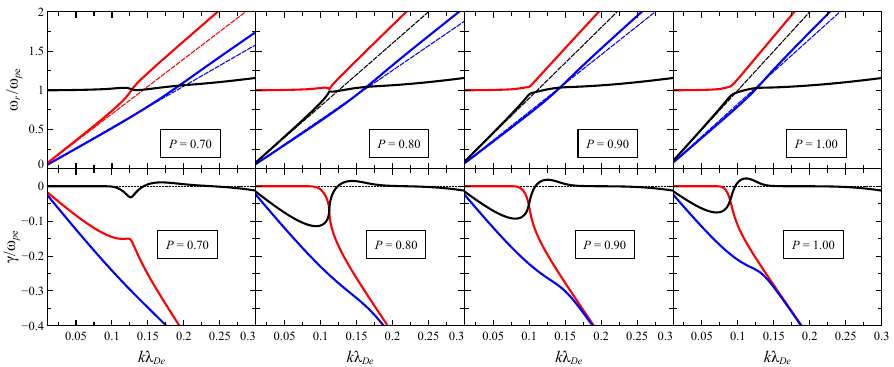}
    \caption{\label{fig:1}Dispersion relations (top panels) and the respective growth/damping rates (bottom panels). For different velocities (left to right) $v_b/v_{te}=7, 8, 9, 10$, and density $n_b/n_0=10^{-3}$. Color codes: black: unstable mode, red: beam mode $\omega^B_1(k)$/modified-Langmuir mode, blue: beam mode $\omega^B_2(k)$\@.
    Dashed lines: approximate dispersion relations $\omega_+(k)$ and $\omega_-(k)$ $(\omega_+ > \omega_-)$.}
\end{figure*}

With this criteria in mind, the solutions of Eq. \eqref{eq:3} are sought in the complex frequency plane and are written as $\omega(k) = \omega_r (k)+ i \gamma(k)$, where $\omega_r(k)$ is the dispersion relation and $\gamma(k)$ is the damping ($\gamma < 0$) or growth ($\gamma > 0$) rate.
Figure \ref{fig:1} shows complex roots of Eq. (\ref{eq:3}) for a beam with fixed density $n_b / n_0 = 10^{-3}$ and temperatures $T_e / T_i = 1$ and $T_b / T_e = 1$, and different velocities,  $v_b / v_{te} =7, 8, 9, 10$\@.
Only the high-frequency solutions (determined by the electronic populations) are shown.

There are always three solutions of the dispersion equation, corresponding to three normal modes of propagation, two of which are always stable ($\gamma (k) <  0$ for all $k$) and one is unstable $(\gamma > 0)$ in a given spectral range of wavenumber values.
One of the solutions, called the \emph{modified-Langmuir} mode, always starts with $\omega_r \approx \omega_{pe}$ at $k = 0$\@.  This mode reduces to the usual Langmuir mode when $v_b = v_e = 0$ and $n_b = 0$\@.
The other two solutions of Eq. (\ref{eq:3}) correspond to the new beam modes that are identified as
$\omega^B_1(k)$ and $\omega^B_2(k)$, with $\omega^B_{r1} > \omega^B_{r2}$ for any $k$\@.
The color codes of the normal modes are as follows. The unstable mode is always depicted by the black curve, the related stable beam or modified-Langmuir mode by red and the other beam mode by blue.
It can be seen that by increasing the beam intensity by increasing its drift velocity, the dispersion relations change and the growing mode no longer follows the modified-Langmuir branch, jumping instead to the beam mode $\omega^B_1$\@.
This transition is visible between the solutions for $P = 0.70$ and $P = 0.80$.

In order to understand the physical origin of the beam modes, we analyze the phase speed of the normal modes, compared with the thermal speed of the different populations.  Denoting by $\xi_\alpha = (\omega_r - k v_\alpha)/kv_{t\alpha}$ the ratio of the phase speed (possibly Doppler shifted by the population's drift) with the thermal speed, a given solution of the dispersion equation is a \emph{fast wave} if $|\xi_\alpha| \gg 1$ or a \emph{slow wave} if $|\xi_\alpha| \ll 1$\@.  For each case, we can employ the following approximations in Eq. (\ref{eq:3}), 
\begin{displaymath}
Z^{\prime}\left(\xi\right) \simeq \begin{cases}
-2\left(1-2\xi^{2}\right), & (|\xi| \ll 1)\\
(\xi^{2}-\nicefrac{3}{2})^{-1}, & (|\xi|\gg 1).
\end{cases}
\end{displaymath}

Both beam modes $\omega^B_1$ and $\omega^B_2$ in Fig. \ref{fig:1} are always fast waves for the ions.  Assuming then that the beam modes are slow waves for the thermal electrons and fast waves for the beam population, one obtains
\begin{equation}
    \left(\omega-kv_{b}\right)^{2}=\frac{\omega_{pb}^{2}+3\left(k^{2}v_{tb}^{2}+\omega_{pe}^{2}v_{tb}^{2}/v_{te}^{2}\right)}{1+\omega_{pe}^{2}/k^{2}v_{te}^{2}},
    \label{eq:BPI1:Beam_modes-2}
\end{equation}
where the effect of the ions was neglected.  Taking the square root on both sides of (\ref{eq:BPI1:Beam_modes-2}), one obtains the dispersion relations $\omega_\pm(k)$, which are the approximations $\omega_+(k) \approx \omega^B_{r1}(k)$ and $\omega_-(k) \approx \omega^B_{r2}(k)$\@.  Actually, the assumption that the beam modes are fast waves for the beam population works better for $\omega^B_{r2}$ than for $\omega^B_{r1}$, and in the latter case the full expression of $\omega_+(k)$ ends up slightly overestimating the real part of the numerical solution\@.
Hence, we will adopt the approximate expressions $\omega_{+}(k) \approx kv_{b}$ and 
\begin{displaymath}
\omega_{-}(k) = kv_{b}-\sqrt{\frac{\omega_{pb}^{2}+3\left(k^{2}v_{tb}^{2}+\omega_{pe}^{2}v_{tb}^{2}/v_{te}^{2}\right)}{\omega_{pe}^{2}+k^{2}v_{te}^{2}}}kv_{te}.
\end{displaymath}
The dispersion relations $\omega_\pm(k)$ are shown as dashed lines in the top panels of Fig. \ref{fig:1}\@.  One can observe that the approximations are very good in the small wavenumber range.

The dispersion relations $\omega_\pm(k)$ provide good approximations for the beam modes in the small wavenumber regime when $P \lesssim 1$, but are in principle valid for any temperature ratio and in the results shown by Fig. \ref{fig:1},  $T_b = T_e$\@.  However, when one of the electronic populations is much hotter than the other, the full expression of $\omega_+(k)$ in (\ref{eq:BPI1:Beam_modes-2}) can be identified with the electron-acoustic mode and electron/electron acoustic instability.\cite{GaryTokar85/08,Gary87/09}  For instance, when the core population is hot and the beam electrons are cold $(T_b \ll T_e)$, the beam modes are definitely fast waves for the beam electrons and $\omega_+(k)$ reproduces Eq. (1) of Ref. \onlinecite{Gary87/09} or the known electron-acoustic relation when $v_b = 0$\@.\cite{GaryTokar85/08}
Therefore, the beam modes $\omega^B_{1,2}(k)$ obtained as numerical solutions of the dispersion equation (\ref{eq:3}), asymptotically reduce to electron-acoustic modes when one population is much hotter than the other.

\begin{figure*}
    \includegraphics[width=\textwidth]{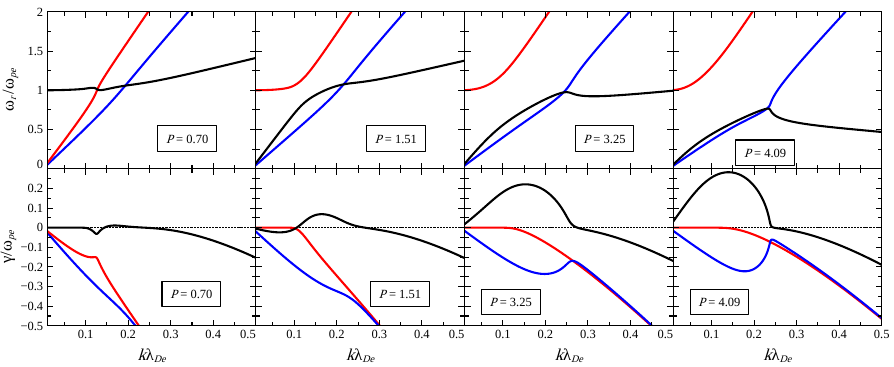}
    \caption{\label{fig:2}Dispersion relations (top panels) and the respective growth/damping rates (bottom panels). For different densities (left to right) $n_b/n_0=10^{-3},10^{-2},10^{-1},2\times 10^{-1}$, and velocity $v_b/v_{te}=7$. Color codes: black: unstable mode, red: beam mode $\omega^B_1(k)$/modified-Langmuir mode, blue: beam mode $\omega^B_2(k)$.
}
\end{figure*}

Figure \ref{fig:2} shows the dispersion curves for a beam with fixed drift velocity $v_b / v_{te} = 7$ and different densities $n_b / n_0 =10^{-3},10^{-2},10^{-1}$, and $2\times10^{-1}$, again with temperatures $T_e / T_i = 1$ and $T_b / T_e = 1$\@. Likewise as shown in Fig. \ref{fig:1}, it can be seen that by increasing the beam intensity, now by increasing its density, the dispersion relations also change, between $P = 0.70$ and $P = 1.51$, in such a way that the growing mode no longer follows the Langmuir mode branch.

For a typical case, with velocity $v_b / v_{te} = 10$ and density $n_b / n_0 = 10^{-3}$, we have $P = 1.00$ and $s = 1.26$. When the beam velocity is increased, with $v_b / v_{te} = 7, 8, 9, 10$,  we obtain the following values for the parameters  $P = 0.70; 0.80; 0.90; 1.00$ and $s = 1.80; 1.57; 1.40; 1.26$. These values obtained for the $s$ and $P$ parameters, which correspond to the transition of the numerical solution, are consistent with the classification made by Cairns \cite{Cairns89/01}.

In essence, for increasing beam intensities, either increasing density or drift velocity, the dispersion relation undergoes a topological change, and the growing branch changes accordingly,
corresponding to the modified-Langmuir mode when $P \lesssim 0.8$ or the $\omega^B_1(k)$ beam mode otherwise.
Therefore, a proper analytical approach would call for an algebraic dispersion relation describing the actual growing mode.

Ideally, a mathematical expression for 
the $\omega^B_1(k)$ mode valid for the whole spectral range of wavenumbers
would be derived rigorously from Eq. \eqref{eq:3}. However, besides being a nontrivial 
task,
the result might not be feasible for the intended analytical calculation. Alternatively, it is conceivable to use a fitting analytical expression for the dispersion relation of the growing mode that captures the key features of the actual numeric solution and still lends itself to a reasonable analytical treatment. Henceforth, this mode is referred to simply as mode B and is defined as
\begin{equation}\label{eq:4}
\omega^{B}(k) \equiv \omega^B_k =\frac{av_{b}k}{1+bv_{b}k} \,,
\end{equation}
where $a$ and $b$ are fitting parameters. %

\begin{figure}
    \includegraphics[width=\columnwidth]{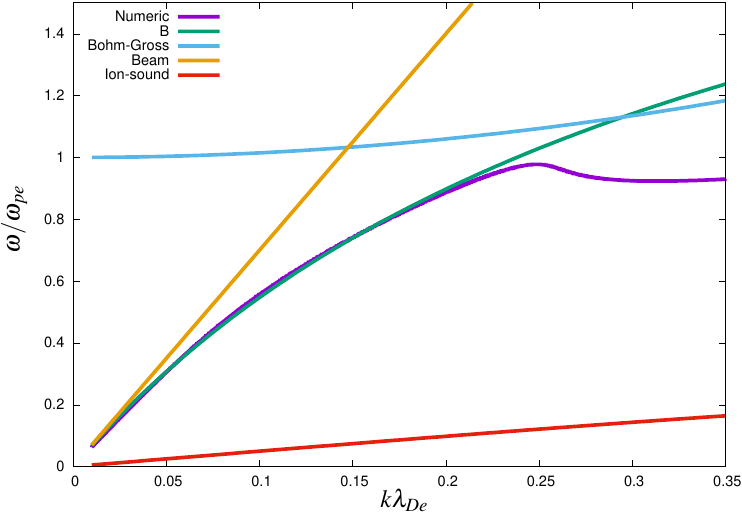}
    \caption{\label{fig:3}Dispersion relations. For $v_b/v_{te}=7$, $n_b/n_0=10^{-1}$, $a=$1 and $b=0.4$ (arbitrary units). For the sake of visualization, the ion-sound mode is out of scale.}
\end{figure}

Figure \ref{fig:3} illustrates the pertinence of using Eq. \eqref{eq:4} to describe the growing mode for high energy beams. Along with the numerical solution and the fitting expression for the growing mode, it is displayed the dispersion relations for Langmuir, ion-sound and beam modes, for comparison.

At this point, it must be emphasized that the model dispersion relation $\omega^B(k)$ is only valid for regions in the parameter space where the $\omega^B_1(k)$ mode is both topologically detached from the modified-Langmuir branch and the unstable mode.  This corresponds to the panels of Figs. \ref{fig:1} and \ref{fig:2} with $P \gtrsim 0.8$.

\section{Numerical analysis}
\label{sec:III}

\subsection{L and S mode approach}

Several studies show results involving beam-plasma systems where the beam is of low intensity and the system is described by the weak turbulence theory. In particular, it is shown that in some cases the quasilinear theory alone satisfactorily describes the evolution of the system even for parameters that are beyond the original limits of the theory.\cite{Gaelzer+03/02, Pavan+10/02, YoonGaelzer02/11, YoonRheeRyu05/06, Yoon05/04}

Nevertheless, the weak turbulence theory allows one to establish the kinetic equations of waves and particles for Langmuir (L) and ion-sound (S) modes, accounting for quasilinear and higher order nonlinear effects. A full account of the theoretical formulation is presented in, e.g., Refs. \onlinecite{YoonGaelzer02/11,YoonRheeRyu05/06,Yoon05/04,Yoon19}\@. Therefore, only the basic equations are shown here. 

\begin{figure}
    \includegraphics[width=\columnwidth]{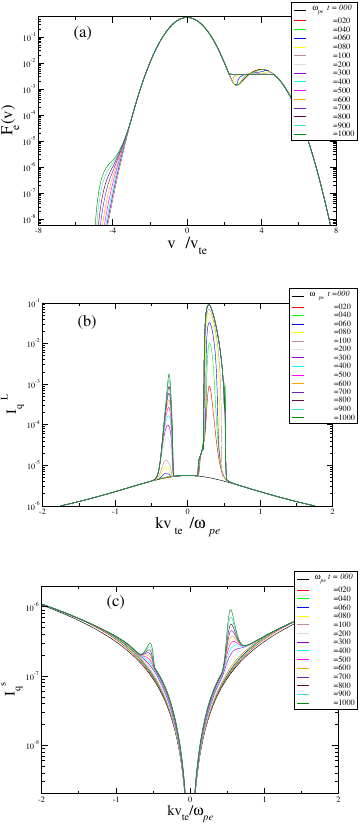}
    \caption{\label{fig:4}Temporal evolution of the beam-plasma system
for several time instants, normalized to the plasma period $\tau = \omega_{pe}t$,
    including quasilinear and 3-wave decay terms, for $v_b/v_{te} = 4$, $n_b/n_e = 10^{-2}$,
$T_b/T_e = 1$ and $T_e/T_i = 7$\@.  (a) Electron distribution function,
normalized as $F_e \rightarrow F_e/v_{te}$, (b) spectral intensity of Langmuir waves and (c) ion-sound waves.}
\end{figure}

When applying the weak turbulence theory formalism to a plasma system, all 
existing normal modes should in principle 
be taken into account to describe the temporal evolution of 
the distribution of plasma particles. %
However, since the spectral intensity of ion acoustic waves is 
usually very small in comparison with the spectral intensity of Langmuir waves, that is, $I^S_k(t) \ll I^L_k(t)$, 
where $I^\alpha_k(t)$ is the spectral intensity of the $\alpha$-th normal mode
(see Figure \ref{fig:4}, for instance), one can make an approximation and neglect the contribution of the former mode when 
analysing the evolution of the distribution function of electrons.
That being said, the kinetic equation for the particle distribution function is given by\cite{Yoon19} %

%
\begin{multline}
    \frac{\partial f_{e}}{\partial t}=\frac{\partial}{\partial v}\sum_{\sigma=\pm1}\frac{\pi e_{e}^{2}}{m_{e}^{2}}\int dk\delta\left(\sigma\omega_{k}^{L}-kv\right)
    \\
    \times\left[\frac{m_{e}}{4\pi^{2}k}\sigma\omega_{k}^{L}f_{e}+I_{k}^{\sigma L}\frac{\partial f_{e}}{\partial v}\right].
\end{multline}
Continuing, the kinetic equations of the wave spectral intensities, neglecting nonlinear scattering terms, are given by\cite{Yoon19}
\begin{multline}\label{di/dt-L}
    \frac{\partial I_{k}^{\sigma L}}{\partial t} = \frac{\pi \omega_{pe}^2}{k^2} \int dv \delta \left( \sigma \omega_{k}^L - kv \right) \left( \frac{ne^2}{\pi}F_e + \sigma \omega_{k}^L I^{\sigma L}_{k} k \frac{\partial F_e}{\partial v} \right) \\
     +  \sum_{\sigma^{\prime},\sigma^{\prime\prime} = \pm 1} \sigma \omega_{k}^{L} \int dk^\prime 
    \frac{\pi e^2}{T_e^2} \frac{\mu_{k - k^{\prime}}}{| k - k^{\prime}|^2} 
         \delta \left( \sigma \omega_{k}^{L} - \sigma^{\prime} \omega_{k^{\prime}}^{L} - \sigma^{\prime \prime} \omega_{k-k^{\prime}}^{S} \right)\\
    \times \left( \sigma \omega_{k}^{L} I_{k^{\prime}}^{\sigma^{\prime}L} I_{k-k^{\prime}}^{\sigma^{\prime\prime}S}\right.
    \left. - \sigma^{\prime} \omega_{k^{\prime}}^{L} I_{k-k^{\prime}}^{\sigma^{\prime\prime}S} I_{k}^{\sigma L} - \sigma^{\prime\prime} \omega_{k-k^{\prime}}^{L} I_{k^{\prime}}^{\sigma^{\prime}L} I_{k}^{\sigma L} \right) ,
\end{multline}
\begin{multline}\label{di/dt-S}
     \frac{\partial I_{k}^{\sigma S}}{\partial t} = \frac{\pi \mu_{k} \omega_{pe}^2}{k^2} \int dv \delta \left( \sigma \omega_{k}^S - kv \right) \\
     \times \left[ \frac{n e^2}{\pi} \left( F_e + F_i \right) 
       + \sigma \omega_{k}^{L} I_{k}^{\sigma S}  k\frac{\partial}{\partial v}  \left( F_e + \frac{m_e}{m_i} F_i \right) \right] \\
      + \sum_{\sigma^{\prime},\sigma^{\prime\prime} = \pm 1} \sigma \omega_{k}^{L} \int dk^{\prime}
         \frac{\pi e^2}{4 T_e^2} \frac{\mu_{k}}{k^2}  
          \delta \left( \sigma \omega_{k}^{S} - \sigma^{\prime} \omega_{k^{\prime}}^{L} - \sigma^{\prime \prime} \omega_{k-k^{\prime}}^{L} \right) \\
     \times \left( \sigma \omega_{k}^{L} I_{k^{\prime}}^{\sigma^{\prime}L} I_{k-k^{\prime}}^{\sigma^{\prime\prime} L} \right. 
      \left. - \sigma^{\prime} \omega_{k^{\prime}}^{L} I_{k-k^{\prime}}^{\sigma^{\prime\prime}L} I_{k}^{\sigma S} - \sigma^{\prime\prime} \omega_{k-k^{\prime}}^{L} I_{k^{\prime}}^{\sigma^{\prime}L} I_{k}^{\sigma S} \right),
\end{multline}
where $\mu_k=k^{3}\lambda_{De}^{3}(m_e/m_i)^{1/2}(1+3T_i/T_e)^{1/2}$\@.
The parameter $\sigma = +1(-1)$ identifies the kinetic equation for the forward- (backward-) propagating wave.

With the equations involving the Langmuir and ion-sound modes, we obtain results for the evolution of particles and waves, with the parameters $n_b/n_e = 10^{-2}$, $v_b/v_{te} = 4$, $T_b/T_e=1$ and $T_e/T_i=7$. The results are shown in terms of normalized wave intensity $I_q^\beta=I_k^\beta/2^{3/2}\lambda^3_{De}n_em_ev_{te}^2$, where $q=kv_{te}/\omega_{pe}$, $\beta$ stands for the wave modes 
$L$, $S$, and $B$,
and the normalized distribution function $f_e=n F_e$.
For consistency with the treatment that consider the single $B$ mode, discussed below in section \ref{subsec:B-mode}, only the decay terms in Eqs. (\ref{di/dt-L}) and (\ref{di/dt-S}) were included.

Figure \ref{fig:4} shows the temporal evolution of the beam-plasma system
according to the traditional approach.  In Fig. \ref{fig:4}(a) it is possible to 
observe
the formation of a superthermal tail in the electron distribution due to absorption of the backscattered mode around $ v_b / v_{te} = -4 $ due to nonlinear effects. 
Fig. \ref{fig:4}(b,c) show the evolution of Langmuir and ion-sound waves. In Fig. \ref{fig:4}(b) it is 
noticeable
an increase in the intensity of the $L$ mode at $kv_{te}/\omega_{pe} \approx v_{te}/v_b$ until the formation of a plateau in the electron distribution, after which ensues the  formation of a secondary wave peak, of the backscattered Langmuir wave, in the region $ kv_{te}/\omega_{pe} \approx - v_{te}/v_b$.
In Fig. \ref{fig:4}(c) it is possible to observe
the formation of a peak of $I^S_q$ at $ kv_{te}/\omega_{pe} \approx 2 v_{te} / v_b$, and after some time the formation of the secondary peak at $ kv_{te}/\omega_{pe} \approx -2 v_{te}/v_b$.

These results are typical for a beam-plasma system and represent a benchmark for comparison with the B mode approach.

\subsection{Wave thermal level}

In order to perform the intended simulations with the B mode,
the initial level of wave intensities is required.
A consistent theory of the evolution of the beam-plasma system, treated as an initial-value problem, demands the starting intensity levels of the plasma eigenmodes.

For a homogeneous, noncollisional plasma in thermal equilibrium, the spectral intensity levels of the wave modes are obtained from the balance between %
(quasi-)thermal
spontaneous emission and induced absorption that results from the condition $\partial I_{\mathbf{k}}^{\alpha}/\partial t \to 0$ (at $t = 0$) in the corresponding wave kinetic equations. For a detailed description, see, \emph{e.g.} Ref. \onlinecite{Yoon19} and references therein.
For the longitudinal modes, the initial levels of the wave spectral intensities are, accordingly, given by\cite{Yoon05/04}

\begin{align*}
    I_{k}^{\sigma L}(t = 0) & = \frac{T_e}{4\pi^2} \frac{1}{1 + 3k^2\lambda_{De}^2},
    \\
    I_{k}^{\sigma S}(t = 0) & =  \frac{T_e}{4\pi^2} k^2\lambda_{De}^2 \left( \frac{1 + k^2\lambda_{De}^2}{1 + 3k^2\lambda_{De}^2} \right)^{1/2} \\
        & \times \frac{\int dv \delta\left( \sigma \omega_{k}^{S} - kv \right) \left( F_e + F_i \right)}{\int dv \delta\left( \sigma \omega_{k}^{S} - kv \right) \left[ F_e + \left( T_e/T_i \right)F_i \right]}.
\end{align*}

These expressions have been adopted in previous works as initial conditions for the beam-plasma problem, when it was assumed that the wave eigenmodes remained the same as in an equilibrium plasma. Nonetheless, not only a beam-plasma system is not in thermal equilibrium, but also the resulting  dispersion relations are topologically different from the traditional wave modes, as can be seen in Figs. \ref{fig:1} and \ref{fig:2}\@.  Consequently, the beam-modified modes are not able to tap on the initial level of fluctuations of Langmuir and ion-sound modes, because they occur along different curves in the $\omega \times k$ diagram.

On the other hand, the remaining regions of the $\omega \times k$ diagram, particularly along the dispersion relations of the beam-modified modes, are filled with the quasi-thermal emission resulting from the electrostatic (and electromagnetic) fluctuations that do not occur along eigenmodes of the plasma, but, instead, are such that the frequencies of the radiation are independent of the wave number.\cite{Yoon19}
Assuming that the plasma is in a state of thermal equilibrium until $t = 0^-$, when the beam is then injected into the system, the intensity levels due to the fluctuations occurring along the new dispersion relations will act as the seed for the corresponding wave kinetic equations.
If $\bigl\langle\delta E_{\parallel}^{2}\bigr\rangle_{\mathbf{k},\omega}$ represents the ensemble average of the fluctuations of the (squared) electrostatic field in Fourier space, \emph{i.e.}, the quasi-thermal electrostatic noise from a plasma at temperature $T$, then\cite{Yoon07,Yoon19}
\begin{equation*}
    \left\langle \delta E_{\parallel}^{2}\right\rangle _{\mathbf{k},\omega}=\frac{T}{2\pi^{3}\omega}\frac{\mathrm{Im}\,\epsilon\left(\mathbf{k},\omega\right)}{\left|\epsilon\left(\mathbf{k},\omega\right)\right|^{2}},
\end{equation*}
where
\begin{equation*}
    \epsilon\left(\mathbf{k},\omega\right)=1+\sum_{a}\frac{2\omega_{pa}^{2}}{k^{2}v_{ta}^{2}}\left[1+\frac{\omega}{kv_{ta}}Z\left(\frac{\omega}{kv_{ta}}\right)\right],
\end{equation*}
is the linear dielectric function of a thermal plasma composed by electrons and ions $(a=e,i)$ and $Z(\xi)$ is the Fried and Conte function.

\begin{figure}
    \includegraphics[width=1\columnwidth]{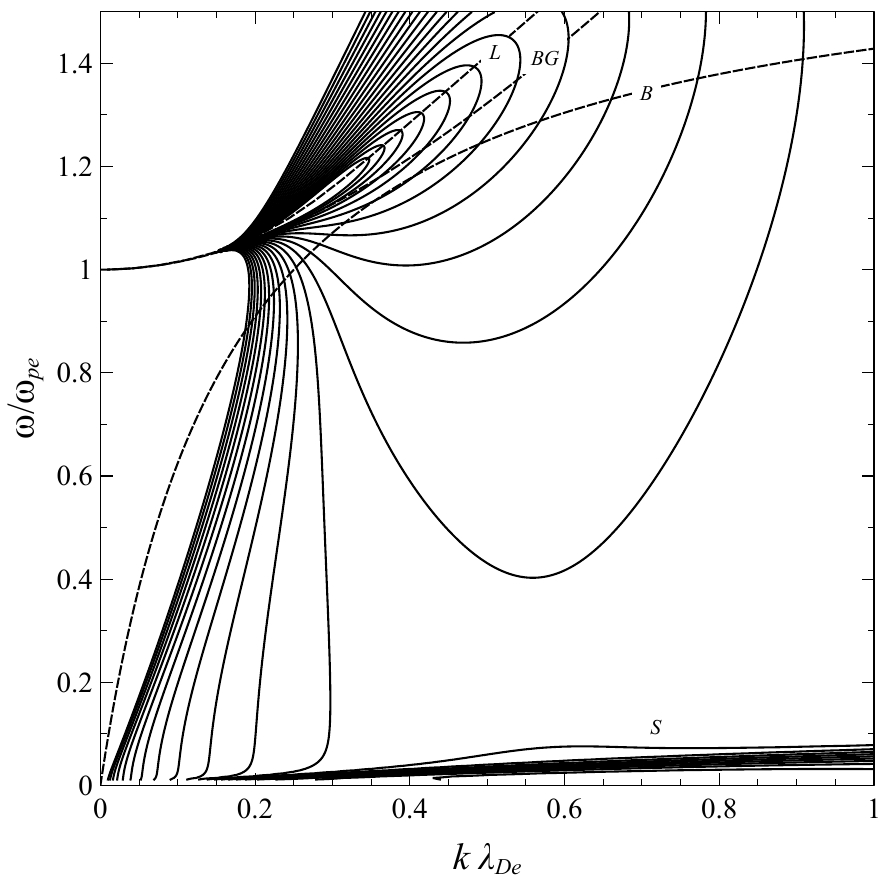}
    \includegraphics[width=1\columnwidth]{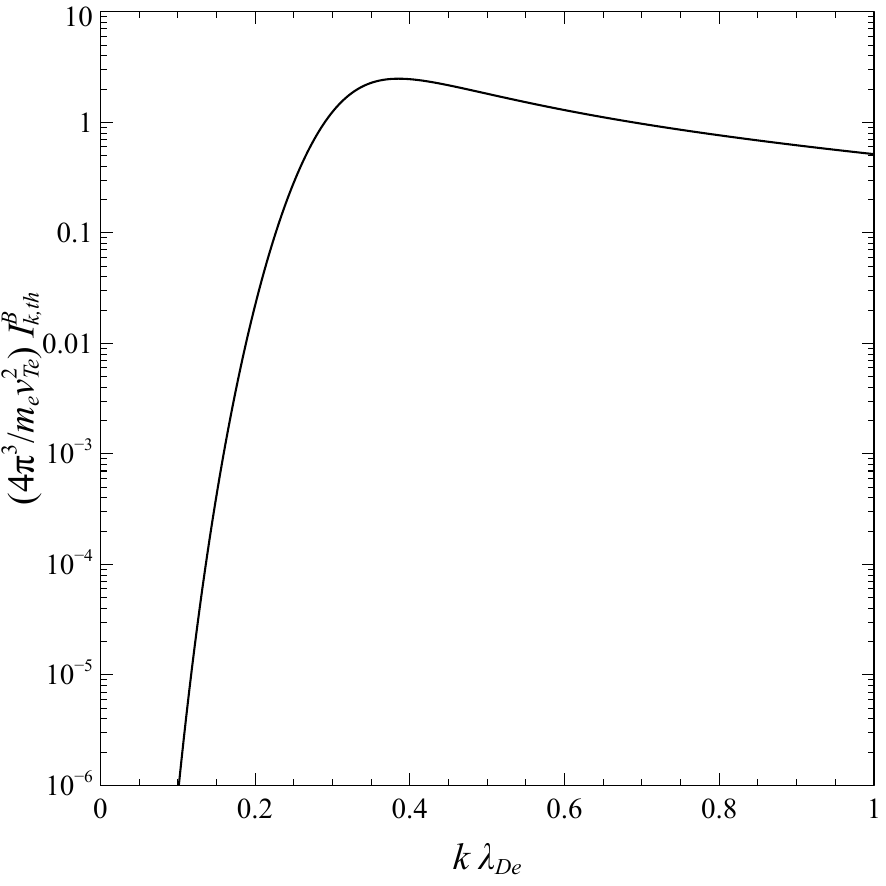}
    \caption{\label{fig:B mode thermal level} Top panel: Contour plots of the quasi-thermal emission $(2\pi^3 \omega_{pe}/T)\bigl\langle\delta E_{\parallel}^{2}\bigr\rangle_{\mathbf{k},\omega}$\@.
    The dashed curves correspond to the dispersion relations of the Langmuir mode $(L)$, Bohm-Gross $(BG)$ and beam mode $(B)$, with $av_b = 10\sqrt{2}v_{te}$ and $b = 0.6\omega_{pe}^{-1}$\@.
    Bottom panel: the normalized thermal spectral intensity along the dispersion relation of the B mode.}
\end{figure}

The top panel of figure \ref{fig:B mode thermal level} shows the electrostatic thermal noise as contour curves.  The contours peak along the dispersion relation of the Langmuir mode, determined by the numerical solution of $\epsilon\left(\mathbf{k},\omega_L\right) = 0$, and shown as the dashed line marked with the label $L$\@.  The contour lines at the bottom correspond to emissions in the ion-sound $(S)$ range.  Also superimposed to the contour lines are the Bohm-Gross dispersion relation
$(BG)$ and the beam mode model $(B)$\@.

The noise level along the dispersion relation of the B mode will act as the initial level of the  spectral intensity of the beam mode.  Since the relation between the electrostatic field fluctuations $\bigl\langle\delta E_{\parallel}^{2}\bigr\rangle_{\mathbf{k},\omega}$ and the spectral intensities of the plasma eigenmodes is given by
\begin{equation*}
\left\langle \delta E^{2}_\parallel \right\rangle _{\mathbf{k},\omega}=\sum_{\sigma=\pm1}\sum_{\alpha}I_{\mathbf{k}}^{\sigma\alpha}\delta\left(\omega-\sigma\omega_{\mathbf{k}}^{\alpha}\right),
\end{equation*}
we can define the quantity $I_{k,\mathrm{th}}^{B} = \omega_{k}^{B}\bigl\langle \delta E^{2}_\parallel \bigr\rangle_{k,\omega_{k}^{B}}$,
interpreted as the “quasi-thermal wave spectral intensity” of the B mode (with dispersion relation $\omega_{k}^{B}$)\@.
Hence,
\begin{equation}
 I_{k,\mathrm{th}}^{B} = I_{k}^{B} (t = 0) =
 \frac{T}{2\pi^{3}}\frac{\mathrm{Im}\,\epsilon\left(k,\omega_{k}^{B}\right)}{\left|\epsilon\left(k,\omega_{k}^{B}\right)\right|^{2}}.
 \label{eq:B-mode_initial-level}
\end{equation}
The bottom panel of Fig. \ref{fig:B mode thermal level} shows the plot of the normalized thermal intensity of the B mode that will be the initial-time level for the kinetic equation of the beam mode.

\subsection{B mode approach}\label{subsec:B-mode}


Mode B 
is incorporated into the modified weak turbulence theory proposed in this work
as follows (more details in the appendix)\@. 
First of all, the B mode is assumed to occur along the dispersion relation (\ref{eq:4}) throughout the whole dynamical evolution of the beam-plasma system.  This is admittedly a shortcoming, since a fully self-consistent treatment would take into account the fact that as the particle distributions and wave intensities evolve, the linear dispersion relations of the normal modes of oscillations will change as well.  However, since we are employing a model dispersion relation for the B mode anyway, we will for now assume that its time evolution can be disregarded as a first approximation.

Likewise for modes L and S, the spectral intensity of mode B is described in terms of the quasilinear and nonlinear contributions, where the nonlinear contribution stands for the 3-wave processes,
\begin{equation*}
    \frac{\partial I_k^{\sigma B}}{\partial t} = \left.\frac{\partial I_k^{\sigma B}}{\partial t}\right|_{QL} + \left.\frac{\partial I_k^{\sigma B}}{\partial t}\right|_{3W} . 
\end{equation*}
The first term on the right side indicates quasilinear interaction terms, and the second term indicates nonlinear terms due to 3-wave interaction.
The kinetic equation for the B mode is numerically solved taking Eq. (\ref{eq:B-mode_initial-level}) as the initial intensity level.

Quasilinear processes for the B mode are ruled by

\begin{multline}\label{eq:7}
\left.\frac{\partial I_{k}^{\sigma B}}{\partial t}\right|_{QL}=\int dv\delta\left(\sigma\omega_{k}^{B}-kv\right) \\
\times \left[ \frac{\left(\sigma\omega_{k}^{B}\right)^{6}}{\omega_{pe}^{4}}\frac{e_{e}^{2}}{k^{2}} f_{e}(v) + \pi n_e\frac{\left(\sigma\omega_{k}^{B}\right)^{3}}{k} I_{k}^{\sigma B} \frac{\partial f_{e}}{\partial v} \right]\,.
\end{multline}

Following the formalism of the weak turbulence theory,\cite{Davidson67/08,CaponiDavidson71/07,Yoon00/12,Yoon05/04} the nonlinear 3-wave decay term is accordingly accounted for as
\begin{align}
\lefteqn{\left.\frac{\partial I_{k}^{\sigma B}}{\partial t}\right|_{3W} = -\pi\frac{\left(\sigma\omega_{k}^{B}\right)^{3}}{\omega_{pe}^{4}}} \hspace{10pt} &
\nonumber \\
& \times
\sum_{\sigma',\sigma''=\pm1}\int dk'
\left|\frac{1}{2}\frac{e}{m_{e}}\frac{\omega_{pe}^{2}}{\sigma\omega_{k}^{B}\sigma'\omega_{k'}^{B}\sigma''\omega_{k-k'}^{B}}\right.
\nonumber \\
& \times\left.\left[\frac{k}{\sigma\omega_{k}^{B}}\text{sign}\left(k-k'\right)+\frac{k'}{\sigma'\omega_{k'}^{B}}\text{sign}\left(k-k'\right)+\frac{\left|k-k'\right|}{\sigma''\omega_{k-k'}^{B}}\right]\right|^{2}
\nonumber \\
& \times\left[\left(\sigma'\omega_{k'}^{B}\right)^{3}I_{k-k'}^{\sigma''B}I_{k}^{\sigma B}+\left(\sigma''\omega_{k-k'}^{B}\right)^{3}I_{k'}^{\sigma'B}I_{k}^{\sigma B}\right.
\nonumber \\
& \left.-\left(\sigma\omega_{k}^{B}\right)^{3}I_{k'}^{\sigma'B}I_{k-k'}^{\sigma''B}\right]\delta\left(\sigma\omega_{k}^{B}-\sigma'\omega_{k'}^{B}-\sigma''\omega_{k-k'}^{B}\right) \,.
\end{align}

%
%
\begin{figure}
    \includegraphics[width=\columnwidth]{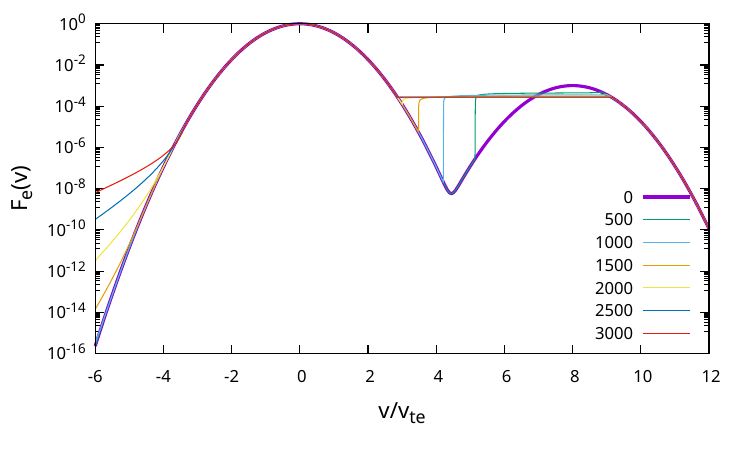}
    \includegraphics[width=\columnwidth]{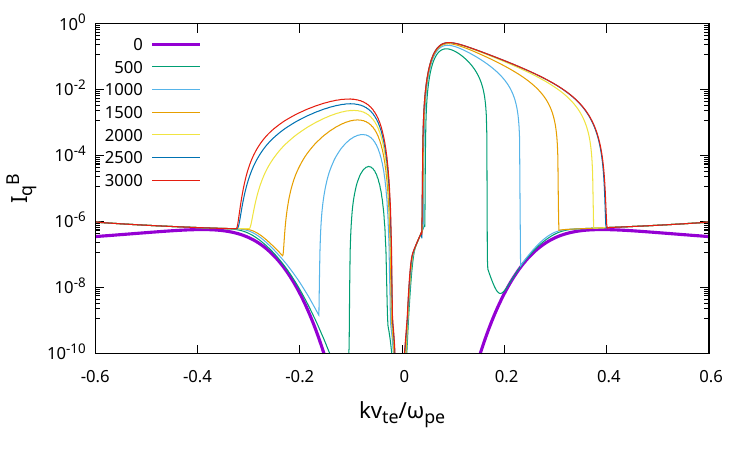}
    \includegraphics[width=\columnwidth]{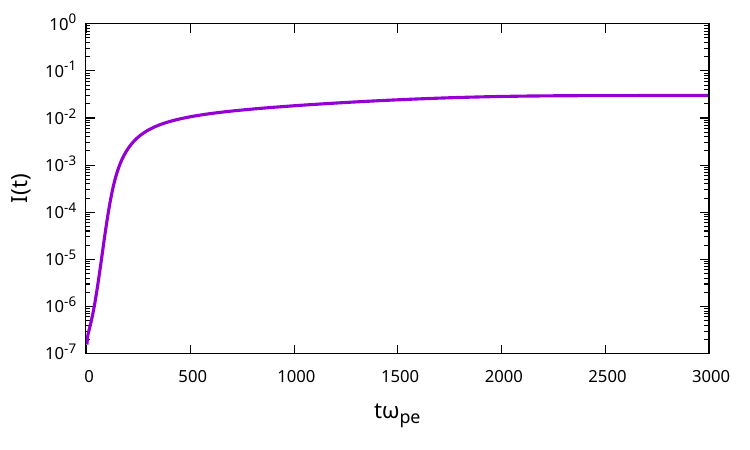}
    \caption{\label{fig:6}Evolution of particles (top panel), spectral intensity of mode B waves (middle panel), including 3-wave decay terms, and temporal evolution of B mode wave energy $I=\int I^B_qdq$ (bottom panel). For velocity $v_b/v_{te}=8$, density $n_b/n_0=10^{-3}$ and $t\omega_{pe}$ from $0$ to $3000$. With the normalization $F_e \rightarrow F_e/v_{te}$.}
\end{figure}

As far as the particles are concerned, we take the simple view that quasilinear approximation is valid. Thus, electrons are ruled by

\begin{multline}\label{eq:6}
\frac{\partial f_{e}}{\partial t}=\frac{\partial}{\partial v}\sum_{\sigma=\pm1}
\frac{\pi e_{e}^{2}}{m_{e}^{2}} \int dk\delta\left(\sigma\omega_{k}^{B}-kv\right)   \\
\times \left[ \frac{m_{e}}{4\pi^{2} k} \frac{\left(\sigma\omega_{k}^{B}\right)^{3}}{\omega_{pe}^{2}} f_{e} +  I_{k}^{\sigma B} \frac{\partial f_{e}}{\partial v} \right]\,.
\end{multline}

Figure \ref{fig:6} shows the evolution of the particles and the spectral intensity of mode B waves with velocity $v_b / v_{te} = 8$ and density $n_b / n_0 = 10^{-3}$, including the 3-wave decay terms. It can be seen in the top panel that particles lose energy to waves until the formation of the plateau and the formation of a superthermal tail due to the absorption of the backscattered mode due to the nonlinear effects. The middle panel shows the evolution of mode B waves, where we see the increase in intensity until the formation of a plateau in the particle distribution, and after that the formation of a secondary peak of backscattered mode B waves. The bottom panel shows that the ratio of B mode wave energy  to particle energy is on the edge of validity of the theoretical approach, which requires this quantity to be small in comparison to unity.

The results of the temporal evolution of waves and particles obtained for high intensity beams show that the behavior of this system is qualitatively similar to the behavior of the system with low intensity beams,
\emph{i.e.}, the traditional weak-turbulence formalism.
This result can be explained by the fact that, although the usual theory considers the coupling of waves involving two distinct modes of oscillation, ion-sound and Langmuir, while the present work considers the existence of a single mode, the dispersion relation of the mode considered, the B mode, has a region of low frequency as well as a region of high frequency, which end up playing the role of the ion-sound and Langmuir waves, respectively.

\section{Final remarks}
\label{sec:IV}

The usual approach, which takes into account nonlinear effects present in the system, is done under the formulation of the weak turbulence theory. In systems where electrostatic oscillations are dominant, this approach considers the interaction between two electrostatic oscillation modes, consisting of the ion-sound mode and the Langmuir (or Bohm-Gross) mode. The dispersion relations that determine these modes are strictly valid for low intensity beams, which for the present study translates into low beam density or low beam drift velocity.

This question was addressed by searching for the numerical solution of the dispersion relation of a beam-plasma system, where the beam is of high intensity. Based on the results obtained from the numerical solution, the dispersion relation was modeled in closed mathematical form in order to allow for algebraic manipulation. The model dispersion relation was used to calculate the coefficients that determine the time evolution of the particle velocity distribution function and the spectral intensity of the waves.

Different densities and drift velocities were considered, representing high intensity beams. The general behavior of waves and particles was compared with those obtained under the weak turbulence theory approach, but calculated considering the interaction between the Langmuir and the ion-sound modes.

Both cases, that is, the usual approach and the approach presently introduced, are different not only by the use of different dispersion relations, which determine the wave-particle and wave-wave interactions, but also by the fact that the number of modes is distinct:  two modes in the usual approach and one mode in the presently introduced approach.

The results show similarities between these two approaches, regarding the behavior of the distribution function and the spectral intensity of the waves. This similarity can be explained by the fact that, although the usual theory considers the coupling of waves involving two different oscillation modes, ion-sound and Langmuir, while the present work considers the existence of a single mode, the dispersion relation of the single mode shows both low and high frequency regions, for small and higher wave number, respectively. These two distinct regions eventually play the role of the ion-sound and Langmuir waves, respectively, and this characteristic is responsible for producing the similarity between the two cases.

The findings contribute to the consistency of weak turbulence theory, suggesting an extension of its validity in the case of beam-plasma systems where the beams are of high intensity. Nevertheless, further developments on the subject would additionally account for the proper temporal growth of wave intensity \cite{Yoon10/11} and the temporal changes in dispersion relations\@.\cite{Pavan+11/04}
%
In particular, the latter aspect is ensued by the transient character of the topological changes in the dispersion relations as the beam undergoes a plateau formation. In contrast, the mode B decay approach assumes that
the mode will remain as a separate branch throughout the entire course of dynamical evolution. In this regard, would be adequate to check the validity of this approach by comparing the theory against a one-dimensional electrostatic particle-in-cell simulation. This might be the subject of a future work. %

\acknowledgments
G.T.I. acknowledges Ph.D. fellowship from Conselho Nacional de Desenvolvimento Cient\'{\i}fico e Tecnol\'ogico (CNPq) -- Brazil\@. 
R.G. acknowledges support provided by CNPq (Brazil), grant no. 313330/2021-2. 

\vspace{0.5cm}

\begin{center}
 ***
\end{center}

\appendix

\section{Fundamental wave and particle equation and resonant delta functions} 

The fundamental wave and particle equations considered are given by\cite{Yoon19}

\begin{widetext}
    \begin{equation}
    \frac{\partial f_{a}}{\partial t}=\frac{\pi e_{a}^{2}}{m_{a}^{2}}\int d\mathbf{k}d\omega\left(\frac{\mathbf{k}}{k}\cdot\frac{\partial}{\partial\mathbf{v}}\right)\delta\left(\omega-\mathbf{k}\cdot\mathbf{v}\right)\left[\text{Im}\frac{m_{a}\epsilon\left(\mathbf{k},\omega\right)}{2\pi^{3}k\left|\epsilon\left(\mathbf{k},\omega\right)\right|^{2}}f_{a}+\left\langle \delta E^{2}\right\rangle _{\mathbf{k},\omega}\left(\frac{\mathbf{k}}{k}\cdot\frac{\partial f_{a}}{\partial\mathbf{v}}\right)\right],
    \end{equation}
    \begin{multline}
    \frac{\partial I_{\mathbf{k}}^{\sigma\alpha}}{\partial t}=-\frac{2\text{Im}\epsilon\left(\mathbf{k},\sigma\omega_{\mathbf{k}}^{\alpha}\right)}{\epsilon'\left(\mathbf{k},\sigma\omega_{\mathbf{k}}^{\alpha}\right)}I_{\mathbf{k}}^{\sigma\alpha}+\sum_{a}\frac{4e_{a}^{2}}{k^{2}\left[\epsilon'\left(\mathbf{k},\sigma\omega_{\mathbf{k}}^{\alpha}\right)\right]^{2}}\int d\mathbf{v}\delta\left(\sigma\omega_{\mathbf{k}}^{\alpha}-\mathbf{k}\cdot\mathbf{v}\right)f_{a}(\mathbf{v})\\
    -\frac{4\pi}{\epsilon'\left(\mathbf{k},\sigma\omega_{\mathbf{k}}^{\alpha}\right)}\sum_{\sigma',\sigma''=\pm1}\sum_{\beta,\gamma}\int d\mathbf{k}'\left|\chi^{(2)}\left(\mathbf{k}',\sigma'\omega_{\mathbf{k}'}^{\beta}\left|\mathbf{k}-\mathbf{k}',\sigma''\omega_{\mathbf{k}-\mathbf{k}'}^{\gamma}\right.\right)\right|^{2}\\
    \times\left(\frac{I_{\mathbf{k}-\mathbf{k}'}^{\sigma''\gamma}I_{\mathbf{k}}^{\sigma\alpha}}{\epsilon'\left(\mathbf{k}',\sigma'\omega_{\mathbf{k}'}^{\beta}\right)}+\frac{I_{\mathbf{k}'}^{\sigma'\beta}I_{\mathbf{k}}^{\sigma\alpha}}{\epsilon'\left(\mathbf{k}-\mathbf{k}',\sigma''\omega_{\mathbf{k}-\mathbf{k}'}^{\gamma}\right)}-\frac{I_{\mathbf{k}'}^{\sigma'\beta}I_{\mathbf{k}-\mathbf{k}'}^{\sigma''\gamma}}{\epsilon'\left(\mathbf{k},\sigma\omega_{\mathbf{k}}^{\alpha}\right)}\right)\delta\left(\sigma\omega_{\mathbf{k}}^{\alpha}-\sigma'\omega_{\mathbf{k}'}^{\beta}-\sigma''\omega_{\mathbf{k}-\mathbf{k}'}^{\gamma}\right).
    \label{eq:waves}
    \end{multline}
\end{widetext}
Where $\epsilon'$ is calculated by taking the long-wavelength limit $(k^2 \rightarrow 0)$. The quantity $\chi_{a}^{(2)}$, the second-order nonlinear susceptibility, can be expressed in the following approximated form,
\begin{align}
    \chi_{a}^{(2)} &(\mathbf{k}^{\prime},\omega^{\prime}|\mathbf{k}-\mathbf{k}^{\prime},\omega-\omega^{\prime}) = - \frac{i}{2} \frac{e_{a}}{m_{a}} \frac{\omega_{pa}^{2}}{\omega \omega^{\prime} (\omega - \omega^{\prime})} \frac{1}{k k^{\prime} |\mathbf{k} - \mathbf{k}^{\prime}|} \nonumber\\ 
    &\times \left[ \frac{k^2}{\omega}\mathbf{k}^{\prime}\cdot(\mathbf{k}-\mathbf{k}^{\prime}) + \frac{k^{\prime 2}}{\omega^{\prime}}\mathbf{k}\cdot(\mathbf{k}-\mathbf{k}^{\prime}) + \frac{(\mathbf{k} - \mathbf{k}^{\prime})^2}{\omega - \omega^{\prime}} \mathbf{k}\cdot\mathbf{k}^{\prime} \right],
    \label{2-order-nonlin-sucep-fast-wave}
    \end{align}
 when the \emph{fast wave condition}\citep{Yoon19} is considered
\begin{equation*}
    \omega^{\prime} \gg k^{\prime}v_{ta}, \qquad \omega - \omega^{\prime} \gg \vert \mathbf{k} - \mathbf{k}^{\prime} \vert v_{ta}, \qquad \omega \gg k v_{ta}.
\end{equation*}

The delta functions characterizing the spontaneous and induced emission terms in the kinetic equation for the beam mode B, along with the diffusion and drift coefficients in the kinetic equation for the particles, can be expressed in a one-dimensional form as follows, with the redefinition $av_{b}\rightarrow v_{b}$  and  $b\rightarrow b/a$ 
\begin{align*}
    \delta \left( \sigma \omega_{k}^{B} - kv \right) &= \left\vert \frac{v_{b}}{\sigma v^{2} - v_{b}v} \right\vert \delta (k - k_{*}), \qquad  k_{*} = \frac{\sigma v_{b} - v}{vbv_{b}}, \\
    \delta \left( \sigma \omega_{k}^{B} - kv \right) &= \frac{1}{|k|} \delta (v - v_{*}), \qquad v_{*} = \frac{\sigma v_{b}}{1 + bv_{b}k}.    
\end{align*}

Expressing the resonance condition among three waves arising from the decay term in a one-dimensional formulation yields

\begin{align}
\lefteqn{\delta(\sigma\omega_{k}^{B}-\sigma^{\prime}\omega_{k^{\prime}}^{B}-\sigma^{\prime\prime}\omega_{k-k^{\prime}}^{B})}\nonumber\\
& \hspace{5pt}=  \left\vert \frac{\sigma^{\prime\prime}v_{b}}{[1+bv_{b}(k-k_{+})]^{2}}-\frac{\sigma^{\prime}v_{b}}{(1+bv_{b}k_{+})^{2}}\right\vert ^{-1}\delta(k^{\prime}-k_{+})\nonumber\\
& \hspace{5pt} +  \left\vert \frac{\sigma^{\prime\prime}v_{b}}{[1+bv_{b}(k-k_{-})]^{2}}-\frac{\sigma^{\prime}v_{b}}{(1+bv_{b}k_{-})^{2}}\right\vert ^{-1}\delta(k^{\prime}-k_{-}).
\label{eq:delta-decay}
\end{align}

The resonant wave vectors $k^{\prime}$ satisfying the delta functions in expression \eqref{eq:delta-decay} are given by
\begin{align}
    k_{+} &= \frac{- \mathcal{B} + \sqrt{\mathcal{B}^2 -4\mathcal{A}\mathcal{C}}}{2\mathcal{A}}, \nonumber \\ 
    k_{-} &= \frac{- \mathcal{B} - \sqrt{\mathcal{B}^2 -4\mathcal{A}\mathcal{C}}}{2\mathcal{A}}, \nonumber 
\end{align}
where
\begin{subequations}
    \begin{equation*}
        \mathcal{B}^2 - 4\mathcal{A}\mathcal{C} \geq 0,
    \end{equation*}
    \begin{equation*}
        \mathcal{A} = b v_{b}^{2} (\sigma^{\prime} + \sigma^{\prime\prime}) - \frac{\sigma b^{2} v_{b}^{3} k}{1 + b v_{b} k},
    \end{equation*}
    \begin{equation*}
        \mathcal{B} = \frac{\sigma b^{2} v_{b}^{3} k^{2}}{1 + b v_{b} k} + v_{b} (\sigma^{\prime\prime} - \sigma^{\prime}) - b v_{b}^{2} k (\sigma^{\prime} + \sigma^{\prime\prime}),
    \end{equation*}
    \begin{equation*}
        \mathcal{C} = v_{b} k (\sigma - \sigma^{\prime\prime}).
    \end{equation*}
\end{subequations}

\bibliographystyle{aipnum4-1}
\bibliography{bibliography}
\end{document}